\documentstyle[prl,aps,twocolumn,floats,epsfig]{revtex}
\begin{document}

\draft

 \renewcommand{\textfraction}{0.0}

\newcommand{\bra}[1]{\langle #1|}
\newcommand{\ket}[1]{|#1\rangle}
\newcommand{\braket}[2]{\langle #1|#2\rangle}


\twocolumn[\columnwidth\textwidth\csname@twocolumnfalse\endcsname

\title{Gamow Shell Model Description of Neutron-Rich Nuclei}

\author{N. Michel,$^1$ W. Nazarewicz,$^{2-4}$ M. P{\l}oszajczak,$^1$ 
and  K. Bennaceur$^{2,5}$}

\address{$^1$GANIL, CEA/DSM-CNRS/IN2P3, BP 5027, F-14076 Caen Cedex 05, France}

\address{$^2$Department of Physics and Astronomy,
              The University of Tennessee,
              Knoxville, Tennessee 37996}
              
\address{$^3$Physics Division,
              Oak Ridge National Laboratory,
              P. O. Box 2008, Oak Ridge, Tennessee 37831}

\address{$^4$Institute of Theoretical Physics,
              Warsaw University,
              ul. Ho\.za 69, PL-00681, Warsaw, Poland}

\address{$^5$ Institut de Physique Nucleaire de Lyon, 
          Universit\'e Claude Bernard - Lyon 1, F-69622 Villeurbanne Cedex, France}

\date{\today}

\maketitle

\addvspace{5mm}
%
%
\begin{abstract}
This work presents the first continuum shell-model study
of weakly bound neutron-rich nuclei involving
 multiconfiguration mixing.
For the single-particle basis, the  complex-energy Berggren ensemble 
representing the bound 
single-particle states,
 narrow resonances, and the  non-resonant continuum
 background  is taken.
Our shell-model Hamiltonian consists of a one-body finite potential and a zero-range
residual two-body interaction. The systems with {\em two} valence neutrons are considered. 
 The Gamow shell model, which is a straightforward extension of the
 traditional shell model,  is shown to be
 an excellent  tool for the microscopic  description of weakly bound systems. 
 It is demonstrated that the residual interaction
 coupling
 to the particle continuum is important; in some cases, 
 it can give rise to the binding of a nucleus.
 
\end{abstract}

\addvspace{5mm}]
\narrowtext

The 
microscopic structure of exotic nuclei near the particle  drip
lines is a topic of great current interest in low-energy nuclear
 physics. Apart from theoretical and experimental nuclear structure
interest,
calculations for nuclei far from stability have astrophysical
implications, especially in the context of   stellar nucleosynthesis.  
What makes this subject
particularly challenging  is the weak binding; hence 
the  closeness of the particle continuum.  

There are many factors which make the coupling to the particle continuum important.
Firstly, even for a bound nucleus, there appears a virtual 
scattering  into the phase space of unbound states. Although this process involves
intermediate scattering states,  the correlated bound states must be particle
stable, i.e., they 
have zero width.
Secondly, the properties of unbound states, i.e., above the particle (or cluster) threshold
 direcly reflect the continuum structure. In addition, continuum coupling directly
  affects   the effective nucleon-nucleon interaction.
  
The treatment of continuum states is an old problem which, in the context
of excited states near or above the decay threshold,
has been  a playground of
the continuum shell model (CSM) \cite{csm}.
In the CSM, including the recently developed
 Shell Model
Embedded in the Continuum (SMEC) \cite{smec},
the scattering 
states  and bound states are treated  on an  equal footing.
So far, most applications of the CSM,  including SMEC, 
have been used to describe situations in which there is coupling to
 one-nucleon decay channels. However, by allowing  only one particle
to be present in  the  continuum,
it  is impossible to apply 
the CSM to     `Borromean systems'
for which $A$- and $(A$-2)-nucleon systems are particle-stable but the intermediate
$(A$-1)-system is  not.
Several approaches, including  the
hyperspherical harmonic method or the coupled-channel approach,
have been developed to study structure and reaction
aspects of three-body weakly bound nuclei \cite{halos}. However,  most of these
models utilize  the particle-core  coupling  which does not allow for the exact treatment
of  core excitations  and the antisymmetrization between the core nucleons and the valence
particles.

The reason for limiting oneself to only one particle in the continuum is two-fold. First, 
 the number of scattering states
needed to properly describe the underlying dynamics can easily go beyond the limit of
what present computers can handle. Second,  treating the
continuum-continuum coupling, which is always present when
two or more particles are  scattered to unbound levels, is difficult.
There have been 
only a few  attempts to treat the multi-particle case \cite{[Wen87]} and, unfortunately,
the proposed numerical schemes, due to their complexity, have never
been adopted in microscopic  calculations involving multiconfiguration mixing.
Consequently,   an entirely
different approach is called for. 
In this work,  we formulate and test the multiconfigurational  shell model 
in the complete Berggren basis.  The resulting Gamow Shell Model (GSM) is
then applied to  systems with two valence neutrons.

The Gamow states (sometimes called Siegert or  resonant states) \cite{Gamow}
 are  generalized eigenstates  of the time-independent 
Schr\"odinger equation with   
complex energy eigenvalues $E=E_0-i\Gamma/2$, where $\Gamma$ stands for
the decay width (which is zero for bound states). These states correspond to
the poles of the $S$-matrix in the complex energy plane  lying on or 
below the positive real axis; they are
regular at the origin  and satisfy
a  purely outgoing  asymptotics. 
In the following, we consider the Gamow states of a one-body spherical
finite potential. The single-particle  (s.p.) basis of
 Gamow states  must be completed by
means of a set of non-resonant continuum states. 
This completeness relation, introduced by Berggren \cite{Berggren}, reads:
\begin{equation}
\sum_{n} \ket{\phi_{nj}} \bra{\tilde{\phi}_{nj}} + \frac{1}{\pi}\int_{L_+}
\ket{\phi_j(k)} \bra{\phi_j(k^*)}  \, dk = 1,
\label{eq10}
\end{equation}
where $\phi_{nj}$ are the Gamow states carrying the s.p. angular momentum $j$,
$n$ stands for all the remaining quantum numbers labeling Gamow states, 
$\phi_j(k)$ are the modified scattering Gamow states, and  
the contour $L_+$ in the complex $k$-plane has to be chosen in such a way that all the
poles 
in the discrete sum in Eq.~(\ref{eq10}) are contained in the domain between $L_+$ and the
real energy axis. 
If $u_{nj}(r)$ stands for the radial part of $\phi_{nj}$, then $\tilde{u}_{nj}(r)={u}_{nj}(r)^*$
and $\tilde{\phi}_{nj}$=${\phi}_{nj}(u\rightarrow \tilde{u})$. 
If the contour $L_+$  is chosen reasonably close to the real energy axis, 
the first term in
(\ref{eq10}) represents the  contribution from bound states and narrow resonances while
the integral part accounts for the non-resonant continuum. 
A number of completeness relations similar to (\ref{eq10})
were studied by Lind \cite{Lind}.

There have been several applications of resonant states  to
problems involving continuum \cite{Vertse}, but in most cases the so-called pole expansion,
  neglecting the contour integral in Eq.~(\ref{eq10}),
was used \cite{nocontour}. The importance of the contour contribution was investigated in 
Refs.~\cite{Ver95,Lind1} in the context of the continuum RPA with separable
particle-hole interactions  where it was concluded
that the non-resonant part
 must be accounted for if one
 aims at a quantitative  description.  
This can be achieved by discretizing the integral in Eq.\ (\ref{eq10}) \cite{contour}:
\begin{equation}
\int_{L_+} \ket{\phi_j(k)} \bra{\phi_j(k^*)} \,
dk= \sum_{i=1}^N \ket{\phi_j(k_i)}
\bra{\phi_j(k^*_i)} \Delta_k,
\label{eq11}
\end{equation}
where $\Delta_k$ depends on the quadrature used (in our case we use the four-point
 interpolation).

In our study,
Gamow states are determined using the generalized shooting method for
bound states
which requires an exterior complex scaling \cite{Vertse}.
The numerical algorithm for finding Gamow states for any finite-depth potential
$U(r)$ has been tested on the example of the
P\H{o}schl-Teller-Ginocchio (PTG)
potential \cite{Gin84},  for which the resonance energies and
wave functions are known analytically. Energies of all PTG resonances with a
width of up to $90$ MeV are reproduced 
with a precision of at least 10$^{-6}$ MeV. 
The antisymmetric two-particle wave functions $\ket{\phi^{(1)}_{i_1} \; \phi^{(2)}_{i_2}}_J$
are obtained in the usual way by coupling the s.p. wave functions of
the considered bound, resonance,  and scattering Gamow states labeled by subscripts $i_1, i_2$
 to the total angular momentum $J$.
The completeness relation for two-particle states,
\begin{equation}
\sum_{i_1,i_2} \ket{\phi^{(1)}_{i_1} \; \phi^{(2)}_{i_2}}_J \,
               {_J} \bra{\phi^{(1)}_{i_1} \; \phi^{(2)}_{i_2}} \simeq 1
\label{eqmn}
\end{equation}                            
can be used to calculate the two-body matrix elements.
 The   radial
integrals entering the Hamiltonian matrix elements were  
regularized separately by an appropriate choice of the angle of
the external complex scaling. The resulting (complex symmetrix) Hamiltonian matrix
can be 
diagonalized using standard methods.

In most applications, one is interested in bound or resonance $N$-body states but not in 
non-resonant continuum. Bound 
states can be clearly identified, because the imaginary part of their energy must be zero. No 
equally simple criterion exists for resonance or scattering states. On the
other hand, the coupling between scattering states and 
resonant states is usually weak, so one can determine the resonances
 using the following two-step procedure. In the first step, 
 the shell-model Hamiltonian is diagonalized in both (i)  the full space including
the contour, and (ii)  the subspace of Gamow states (pole expansion). 
In the second step,
one identifies the
eigenstates of  (i)  which have the largest overlap
with those of the second  diagonalization. 
For the case of two valence particles discussed in this work, one can 
include in the basis up to 50 states in the non-resonant scattering continuum. 
For greater  dimensions, e.g.,  for a larger number of valence particles, 
this method becomes impractical and the
perturbative correction methods  must be used \cite{future}.

In the following  exploratory GSM calculations, we shall consider
two  cases: (i) $^{18}$O with the inert $^{16}$O core and two active 
neutrons  in the $sd$ shell, and (ii) $^{6}$He with the inert
$^{4}$He core and two active neutrons  in the $p$ shell. 
 Our aim is not to give the
 precise description of $^{18}$O and $^{6}$He (for this, one would need
 a realistic Hamiltonian and a larger configuration space), but rather
 to illustrate the method, its basic ingredients, and underlying features.

\begin{center}
{\it The} ``$^{18}$O" {\it case}
\end{center}
The s.p.\ basis was generated by a Woods-Saxon (WS)
 potential with 
the radius $R_0$=3.05\,fm, the surface diffuseness $d$=0.65\,fm,
the potential depth $U_0$=--55.8\,MeV, and the strength of the spin-orbit term 
$U_{so}$=6.06\,MeV. 
With  this choice of parameters, the single particle 
$0d_{5/2}$ and $1s_{1/2}$ states are bound, with s.p.\ energies -4.14\,MeV,  
and -3.27\,MeV, respectively, and
$0d_{3/2}$ is a resonance with the s.p.\ energy 0.9--i0.97\,MeV. Energies of
these s.p. states are close to the s.p. states of $^{17}$O.

 The completeness relation requires taking
the 
$s_{1/2}$, $d_{5/2}$, and $d_{3/2}$ non-resonant continuums.
For the $1s_{1/2}$ and $0d_{5/2}$ bound  states,  their
 non-resonant continuums can be chosen
along the real momentum axis. 
Since, to the first order, the inclusion of these
 continuums should only result in the renormalization of the effective interaction,
they are ignored
for the purpose of the present exercise whose main focus is the neutron emission.
On the contrary, $0d_{3/2}$ is a
resonance state, so the associated contour  has to be complex to produce the correct
energy width. 
The contour $L_+$  representing  the $d_{3/2}$-continuum was chosen to consist of 
three straight segments
connecting the points
$k_1$=0--i0, $k_2$=0.3--0.2i, $k_3$=0.5--i0, and $k_4$=2.0--i0 (all in fm$^{-1}$).
The  strength of the $\delta$-force was taken to be $V_0$=-350\,MeV\,fm$^3$.

The completeness of the Gamow  basis depends on
the number of discretized scattering basis states considered. 
Table~\ref{TableO} illustrates this dependence.
The real part of energy represents the binding energy  of a state
with respect to the $^{16}$O core, i.e., the two-neutron separation energy.
For the resonance states, 
the real and imaginary parts of energy
 do not change much by increasing the  number  of  scattering  states.
 On the other hand, bound states acquire a very small {\em negative}
 width which does not exceed several keV. This
spurious negative width  depends strongly
on the basis size, and the convergence to zero is both slow and non-monotonic. 
The presence of a small and  negative width is a feature of particle-bound
states obtained  in the GSM. The results displayed 
in Table~\ref{TableO} show that only about 10-20 vectors in the scattering
continuum are sufficient to keep the error of calculated energies and widths
at the acceptable level. It is also clear that  the ``no-pole"
approximation (inclusion of no  scattering states) gives a rather poor description of 
bound and near-threshold states
while it works fairly well for high-lying states carrying a sizeable width. In this respect,
this result is consistent with the conclusions of Refs.~\cite{Ver95,Lind1}.

In the considered example,  the
calculated one-neutron threshold is -4.142 MeV ($0d_{5/2}$ energy) while the
two-neutron threshold is at zero (the binding energy of the core). Consequently,
few states shown in Table~\ref{TableO} are two-neutron bound but 
one-neutron unstable, {\it e.g.} the calculated width, 57\,keV,
of the $J^{\pi}$=$2_4^{+}$ state at $E_0$=-2.61\,MeV, 
characterizes single neutron emission from this state. The higher-lying states
shown in Table~\ref{TableO} are unstable with respect to both one- and two-neutron emission.

\begin{table}[ht]
\caption{Dependence of  energies (left number, in MeV) and neutron widths (right number, 
in keV) of calculated
states in  $^{18}$O on the number of discretized scattering basis states along the contour $L_+$.
}
\label{TableO}
\begin{center}
\begin{tabular}{c|cccc}
$J_{\pi}$    & 0 states & 10 states & 30 states & 50 states \\
\hline
 & \multicolumn{4}{c} {Bound States}\\
$0_1^{+}$ & {-11.73,~-131} & {-12.11,~-2.91}  & {-12.12,~0.27}  & {-12.12,~0.21}  \\
$2_1^{+}$ & {-9.20,~-26.37} & {-9.24,~-0.51} & {-9.24,~-0.031}  & {-9.24,~-0.032}  \\
$4_1^{+}$ & {-8.64,~-13.51} & {-8.64,~-0.25} & {-8.64,~-0.004}  & {-8.64,~-4E-4}  \\
$0_2^{+}$ & {-7.66,~-1.08} & {-7.66,~-0.324} & {-7.66,~-0.264} & {-7.66,~-0.260}  \\
$2_2^{+}$ & {-7.85,~-4.64} & {-7.86,~-0.167} & {-7.86,~-0.066}  & {-7.86,~-0.049}  \\
& \multicolumn{4}{c} {Resonances}\\ 
$4_2^{+}$ & {-4.00,~-89.8} & {-4.03,~-8.15} & {-4.04,~0.54} & {-4.04,~0.54}   \\
$2_3^{+}$ & {-3.48,~29.08} & {-3.48,~43.08} & {-3.49,~44.65} & {-3.49,~44.65} \\
$2_4^{+}$ & {-2.60,~46.45} & {-2.61,~55.82} & {-2.61,~57.02} & {-2.61,~57.04}  \\
$0_3^{+}$ & {1.05,~-172} & {0.94,~-18.64} & {0.94,~0.82} & {0.94,~0.76}  \\
$2_5^{+}$ & {1.63,~112.4} & {1.64,~111.9} & {1.63,~117.5} & {1.63,~118.5}  \\
\end{tabular}
\end{center}
\end{table}

In order to illustrate the configuration mixing induced by the two-body interaction in the GSM,
Table~\ref{Table2O} shows the complex squared shell-model amplitudes
calculated for  the bound ($0^+_1$ and  $2^+_1$) and resonance  ($2^+_3$)  states in $^{18}$O.
All eigenstates are normalized according to  Berggren \cite{Berggren,Lind}:
$\sum_n{c_n^2} = 1$. One should notice  that -- contrary to the conventional
 SM case --  no modulus square appears
in the normalization. 
This   implies that the probabilistic interpretation of $c_n$ 
must be generalized  \cite{Berggren,Berggren1}, i.e, 
when computing  expectation values  
the real part  of $c_n^2$ should be  associated with the mean value 
while   the imaginary part represents the uncertainty 
due to the decaying nature of the state.
As seen in  Table~\ref{Table2O},  the contribution from the non-resonant  continuum 
plays a different role compared to  that from the  resonant states. 
Firstly,  it is generally smaller than 
the leading components involving resonant orbits, though in the example shown in
Table~\ref{Table2O} the contribution of the $0d_{3/2}^2$ resonance   is similar in magnitude.
 Secondly, according to our calculations,
 the inclusion of the contour primarily
 affects  the imaginary part. Finally, 
 the contribution from two particles in the non-resonant
continuum, $L_+^{(2)}$, even though smaller than the one-particle contribution,
 $L_+^{(1)}$, is  not negligible.

\begin{table}[ht]
\caption{Squared amplitudes of different configurations in $0_1^+$, 
$2_1^+$,  and $2_3^+$   states of $^{18}$O. The sum of squared amplitudes
of all Slater determinants including one and two particles in the non-resonant
continuum are denoted by $L_+^{(1)}$ and $L_+^{(2)}$, respectively.
50 discretized scattering states were used.}
\label{Table2O}
\begin{center}
\begin{tabular}{c|ccc}
$c^2$ & $0^+_1$ & $2^+_1$ & $2^+_3$ \\
\hline
$1s_{1/2}^2$ & {0.05--i9.1E-6} & {--} & {--}  \\
$0d_{5/2}^2$ & {0.91--i6.1E-6} & {0.86+i1.2E-5} & {6.9E-3--i5.3E04}  \\
$0d_{3/2}^2$ & {0.02--i5.3E-3} & {1.9E-3--i4.4E-4} & {1.6E-3--i5.2E-4}\\  
$1s_{1/2}0d_{5/2}$ & {--} & {0.13--i1.2E-5} & {4.5E-3--i3.6E-4} \\  
$1s_{1/2}0d_{3/2}$ & {--} & {4.6E-3--i5E-4} & {0.03--i4.9E-3} \\  
$0d_{5/2}0d_{3/2}$ & {--} & {7.7E-3--i8.4E-4} & {0.96+i4.5E-3} \\  
$L_+^{(1)}$ & {1.3E-2+i3.8E-3} & {2.5E-3+i1.7E-3} & {-1.2E-3+i1.7E-3}\\  
$L_+^{(2)}$ & {3.4E-3+i1.5E-3} & {1.6E-4+i9.9E-5} & {5.5E-5+i4.3E-5} 
\end{tabular}
\end{center}
\end{table}

\begin{center}
{\it The} ``$^{6}$He" {\it case}
\end{center}
A description  of the Borromean nucleus $^{6}$He is a challenge for the GSM. $^{4}$He is a 
well-bound system with the one-neutron emission threshold at 20.58\,MeV. On the
contrary, the nucleus $^{5}$He,  with
one neutron in the $p$ shell,  is unstable with respect to the neutron emission. Indeed,
the $J^{\pi}=3/2_1^{-}$
ground state of $^{5}$He lies 890\,keV above the neutron emission
threshold and its neutron width is large, $\Gamma$=600\,keV. The first excited state, 
$1/2_1^{-}$, is a very broad resonance ($\Gamma$=4\,MeV) that lies
4.89\,MeV above the  threshold. $^{6}$He, on the contrary, is bound with 
the two-neutron emission threshold at 0.98\,MeV and one-neutron emission
threshold at 1.87\,MeV. The first excited state $2_1^+$ at 1.8\,MeV in 
$^{6}$He is neutron unstable with a width $\Gamma$=113\,keV. 
In our GSM calculations, the states in $^{5}$He
are viewed  as one-neutron resonances outside of the $^{4}$He core.
A good fit to  $3/2_1^{-}$ and $1/2_1^{-}$ states in $^{5}$He is obtained by taking the WS
potential with  $R_0$ = 2.0 fm, 
$d$=0.65\,fm,  $U_0$ = -47.0 MeV, and  $U_{so}$ = 7.5 MeV. With this potential, one finds the 
single-neutron resonances $p_{3/2}$ 
and $p_{1/2}$ at  $E$=0.745--i0.32 MeV  and $E$=2.13--i2.94 MeV, respectively.
The s.p.\ basis has been  restricted to the $0p_{3/2}$ resonance state
and the $p_{3/2}$ non-resonant  continuum. 
[The $0p_{1/2}$ resonance is very broad and cannot be included in a meaningful way
in the discrete sum in Eqs.~(\ref{eq10},\ref{eq11}). Consequently, 
following the reasoning applied to the $^{18}$O case,
 the $p_{1/2}$ contour  along  the real $k$-axis has been ignored.]
The $L_+$-contour for the
non-resonant $p_{3/2}$ continuum is chosen to enclose 
the $0p_{3/2}$ resonance:
$k_1$ = 0--i0,
$k_2$ = 0.2--0.2i, $k_3$=0.5--i0,  and $k_4$=2.0--i0 fm$^{-1}$. 
The  strength of the $\delta$-force was taken to be $V_0$=650\,MeV\,fm$^3$.
The number of points used to discretize  the scattering continuum is 50, though even with 15
points  the results are reasonably stable. With this precision, 
we reproduce the most important feature of $^6$He: {\em the ground state
is particle-bound, despite the fact that all the basis states lie in the continuum.}
Table~\ref{table4}  shows the structure of wave functions 
of $0_1^+$ and $2_1^+$ states in $^6$He. The important
contribution from the non-resonant continuum is seen, 
even for the $0_1^+$ ground state  which is  particle stable. 
In spite of a very crude Hamiltonian,
the neglect of the exact three-body asymptotics, etc.,  the calculated
 ground state energy  $E$=-0.951--i0.01\,MeV reproduces surprisingly well the
experimental ground state energy with respect to the two-neutron emission
threshold. The excited state $2_1^{+}$ is predicted to lie at 2.25 MeV,
slightly above the experimental value, and its width 
$\Gamma$=700\,keV which depends sensitively on the position of the state with
respect to the emission threshold is somewhat too high as well.

\begin{table}[ht]
\caption{Same as in Table{\ }\protect\ref{Table2O}, except for 
 $0_1^+$ and $2_1^+$ states in  $^{6}$He.}
\label{table4}
\begin{center}
\begin{tabular}{c|cc}
$c^2$ & $0^+_1$ & $2^+_1$ \\
\hline
$0p_{3/2}^2$ & {0.95--i0.79} & {1.011+i0.0044}   \\
$L_+^{(1)}$ & {0.11+i0.76} & {-0.011--i0.0049}  \\  
$L_+^{(2)}$ & {-0.06+i0.03} & {-1.3E-4+i4.8E-4}  
\end{tabular}
\end{center}
\end{table}

In conclusion, the  complex-energy Berggren ensemble 
 is applied for the first time
in shell-model calculations for
two-neutron  states near the particle-emission threshold. 
The results are very encouraging.
It is seen
that the contribution from the non-resonant continuum is important, 
especially for bound and near-threshold states. The particle-bound
states calculated in the GSM are characterized by 
small and negative widths which show non-monotonic behavior as a function
of the basis size. According to our experience, only about 10--20 vectors in the scattering
continuum are sufficient to keep the error of calculated energies and widths
at an acceptable level. 
With a simple interaction, such as the $\delta$-force, we calculated
the low-lying states of $^{18}$O and $^{6}$He and discussed their properties with respect to
neutron emission. 
Last, but not least, pairing correlations due to the continuum-continuum scattering
have been shown to bind the ground
state of $^{6}$He with a completely unbound basis provided by the s.p.
resonances of $^5$He. Further applications of the GSM are in progress \cite{future}. 

%
%
\acknowledgements
We wish to thank J. Oko{\l}owicz and T. Vertse for useful discussions and suggestions.
This work was supported in part by the U.S.\ Department of Energy
under Contract Nos.\ DE-FG02-96ER40963 (University of Tennessee) and
 DE-AC05-00OR22725
with UT-Battelle, LLC (Oak Ridge National Laboratory).

\end{document}